\def\draftversion{false}

\RequirePackage{ifthen}
\ifthenelse{\equal{\draftversion}{true}}{
  \documentclass[aps,prb,galley,superscriptaddress,
       longbibliography]{revtex4}
}{
  \documentclass[aps,prb,reprint,superscriptaddress,
       longbibliography]{revtex4-1}
}

\usepackage[colorlinks=true,linkcolor=blue,citecolor=blue,urlcolor=blue]{hyperref}

\usepackage{soul}  

\usepackage{hyperref}
\usepackage{setspace} 
\usepackage{graphicx}
\usepackage{amsmath}
\usepackage{amsmath}
\usepackage{amssymb}
\usepackage{verbatim}
\usepackage{latexsym}
\usepackage{enumerate} 
\usepackage{bm} 
\usepackage{multirow}

\usepackage[usenames,dvipsnames]{color}

\ifthenelse{\equal{\draftversion}{true}}{
  \marginparwidth 2.7in
  \marginparsep 0.5in
  \newcounter{comm} 
  \def\commnext{\stepcounter{comm}}
  \def\commtext{{\bf\color{blue}[\arabic{comm}]}}
  \def\commmar{{\bf\color{blue}[\arabic{comm}]}}
  \def\dvm#1{\commnext\marginpar{\small DV\commmar: #1}\commtext}
  \def\tmm#1{\commnext\marginpar{\small TM\commmar: #1}\commtext}
  \def\mlab#1{\marginpar{\small\bf #1}}
  
}{
  \def\dvm#1{}
  \def\tmm#1{}
  \def\mlab#1{}
  
}

\begin{document}

\title{Temperature dependence of the bulk Rashba splitting in
the bismuth tellurohalides}

\author{Bartomeu Monserrat}
\email{bm418@cam.ac.uk}
\affiliation{Department of Physics and Astronomy, Rutgers
University, Piscataway, New Jersey 08854-8019, USA}
\affiliation{TCM Group, Cavendish Laboratory, University of
Cambridge, J.\ J.\ Thomson Avenue, Cambridge CB3 0HE, United
Kingdom}
\author{David Vanderbilt}
\affiliation{Department of Physics and Astronomy, Rutgers
University, Piscataway, New Jersey 08854-8019, USA}

\date{\today}

\begin{abstract}
We study the temperature dependence of the Rashba-split bands
in the bismuth tellurohalides BiTe$X$ $(X=$ I, Br, Cl) from
first principles. We find that increasing temperature reduces the
Rashba splitting, with the largest effect observed in BiTeI with
a reduction of the Rashba parameter of $40$\% when temperature
increases from $0$\,K to $300$\,K.
These results highlight the inadequacy of previous
interpretations of the observed Rashba splitting in terms of
static-lattice calculations alone.
Notably, we find the opposite trend, a strengthening of the Rashba
splitting with rising temperature, in the pressure-stabilized
topological-insulator phase of BiTeI.
We propose that the opposite trends with temperature
on either side of the topological phase
transition could be an experimental signature for identifying it.
The predicted temperature dependence is consistent with optical
conductivity measurements, and should also be observable using
photoemission spectroscopy, which could provide further insights into 
the nature of spin splitting and topology in the bismuth tellurohalides.
\end{abstract}

\maketitle

\section{Introduction}

The spin-orbit interaction, which arises in the nonrelativistic
limit of the Dirac equation, is an interaction between the
spin and orbital degrees of freedom of electrons in solids that
drives a number of phenomena, such as the spin splitting of
bands~\cite{winkler_book} and the band inversion in topological
insulators.~\cite{topological_insulators_review,topological_insulators_review2}

In systems that break inversion symmetry and have a 
symmetry axis along $\mathbf{e}_3$, the spin-orbit interaction 
can be characterized by the two-band 
Rashba Hamiltonian~\cite{rashba_original_paper,rashba_second_paper,winkler_book}
\begin{equation}
H_{\mathrm{R}}=\alpha_{\mathrm{R}}\bm{\sigma\cdot\mathbf{k}\times e}_3, \label{eq:rashba}
\end{equation}
where $\alpha_{\mathrm{R}}$ is the Rashba parameter which captures the
strength of the spin-orbit interaction,
$\bm{\sigma}=(\sigma_1,\sigma_2,\sigma_3)$ are the Pauli matrices, and
$\mathbf{k}$ is the $3$-dimensional crystal momentum.
The Rashba Hamiltonian leads to a spin splitting of the electron bands 
in the plane perpendicular to $\mathbf{e}_3$, characterized by the Rashba
momentum $k_{\mathrm{R}}$ by which bands of opposite spin polarization
are shifted, and the Rashba energy $E_{\mathrm{R}}$ at that
band minimum. In terms of these parameters,
$\alpha_{\mathrm{R}}=2E_{\mathrm{R}}/k_{\mathrm{R}}$.

The spin splitting in energy and momentum provided by the Rashba
spin-orbit interaction drives a number of interesting phenomena, such as
the spin Hall effect,~\cite{spin_hall_effect} the Edelstein
effect,~\cite{edelstein_effect} the spin galvanic
effect,~\cite{spin_galvanic_effect} and superconductivity with
non-centrosymmetric pairing.~\cite{noncentrosymmetric_superconductivity}
This rich phenomenology can be exploited in spintronic devices, in which
external electric and magnetic fields, as well as light, are used to
control the spin degrees of freedom.~\cite{spintronics_review} Due to
this fundamental and applied interest, a significant research effort is
dedicated to the discovery and characterization of materials exhibiting
strong Rashba splitting. Rashba splittings have been measured at
surfaces,~\cite{rashba_surface_ag_on_bi}
interfaces,~\cite{rashba_interface_ingaas_inalas} and most recently a
giant Rashba effect has been discovered in the bulk and surfaces of the
polar bismuth tellurohalides BiTe$X$ ($X=$ I, Br,
Cl).~\cite{bitei_giant_rashba,grioni_rashba_bitei_valence_conduction,
   bitei_disentanglement_bulk_surface,bitebr_bitecl_rashba_experiment}
We direct the readers to a recent review that highlights the quantum
phenomena associated with the spin-orbit physics of the bismuth
tellurohalides.~\cite{bahramy_bitei_review}

First-principles methods have been used to interpret experimental
measurements of the Rashba splitting in the bismuth tellurohalides and
to understand the microscopic origin of the
effect.~\cite{bitei_giant_rashba,bitebr_bitecl_rashba_experiment,
  grioni_rashba_bitei_valence_conduction,
  bitei_disentanglement_bulk_surface,bahramy_bitei_review,
  bahramy_soc_microscopic_bitei,chulkov_surface_bitex}
With this understanding, it becomes possible to use first principles
calculations to search for other materials exhibiting large bulk Rashba
effects, an effort that has led to the proposal and discovery of a
number of promising
candidates.~\cite{picozzi_gete_rashba,kim_rashba_halide_perovskite,
  narayan_abc_feti,picozzi_gete_rashba_experiment}
Most first-principles calculations in this context are based on
semilocal approximations to density functional theory (DFT), although
some recent work has highlighted that many-body electron correlations
can have large effects on the calculated Rashba
parameters.~\cite{bitei_many_body} These calculations are invariably
performed within the static lattice approximation, which for materials
containing heavy elements with small quantum zero-point energies is
approximately equivalent to zero temperature. This could make comparison
with experiment problematic, as measurements are typically performed at
finite temperatures. Furthermore, as the effects of temperature on
Rashba splitting are at present poorly known, this also raises the 
question of the relevance of first-principles predictions of novel 
Rashba materials for spintronic applications, as devices would be 
required to operate at room temperature.

In this work we use first principles methods to study the role of
temperature on the Rashba-split electronic states. The temperature
dependence of electronic states has two origins: electron-phonon
coupling and thermal expansion. We explore both contributions to the
bulk Rashba splitting of the bismuth tellurohalides, and find that their
contributions are comparable. We find that the Rashba parameter
decreases with increasing temperature, with the strongest change
observed in BiTeI for which $\alpha_{\mathrm{R}}$ decreases by $40$\%
in the conduction bands from $0$\,K to $300$\,K. We also predict the
opposite trend, an enhancement of Rashba splitting with increasing
temperature, in the topological-insulator phase of BiTeI, which is
stabilized under hydrostatic pressure. This result suggests that
monitoring the sign of the band gap change with temperature could be a
useful signature for identifying a topological phase transition. Our
results imply that quantitative first-principles predictions of Rashba
splitting cannot neglect the effects of temperature.

The rest of the paper is organized as follows. In Sec.~\ref{sec:comp} we
describe the theoretical formalism used for the description of
temperature, and provide the computational details of the
first-principles calculations. We present the temperature dependence of
the Rashba splitting of the bulk bismuth tellurohalides in
Sec.~\ref{subsec:bulk_bitex}. In Sec.~\ref{subsec:ti} we describe our
results for BiTeI in the topological-insulator phase. We discuss the
consequences of our results for theory and experiment in
Sec.~\ref{subsec:exp}, and summarize our findings in
Sec.~\ref{sec:summary}.

\section{Computational details}  \label{sec:comp_details}
\label{sec:comp}

\subsection{First principles calculations}
\label{subsec:first_principles}

We perform first-principles calculations based on DFT using {\sc
vasp}~\cite{vasp1,vasp2,vasp3,vasp4} and the projector augmented-wave
method.~\cite{paw_original,paw_us_relation}
We choose an energy cut-off of $500$\,eV and a $\mathbf{k}$-point grid of
size $8\times8\times8$ for BiTeI and BiTeBr, and $8\times8\times4$ for
BiTeCl. For the supercell calculations we use commensurate grids. All
calculations are performed including the spin-orbit interaction with the
second variational method,~\cite{soc_second_variation_method} in which 
the spin-orbit interaction is included as a perturbation to the scalar
relativistic Hamiltonian.

\subsection{Structural properties}
\label{subsec:structures}

\begin{figure}
\centering
\includegraphics[scale=0.11]{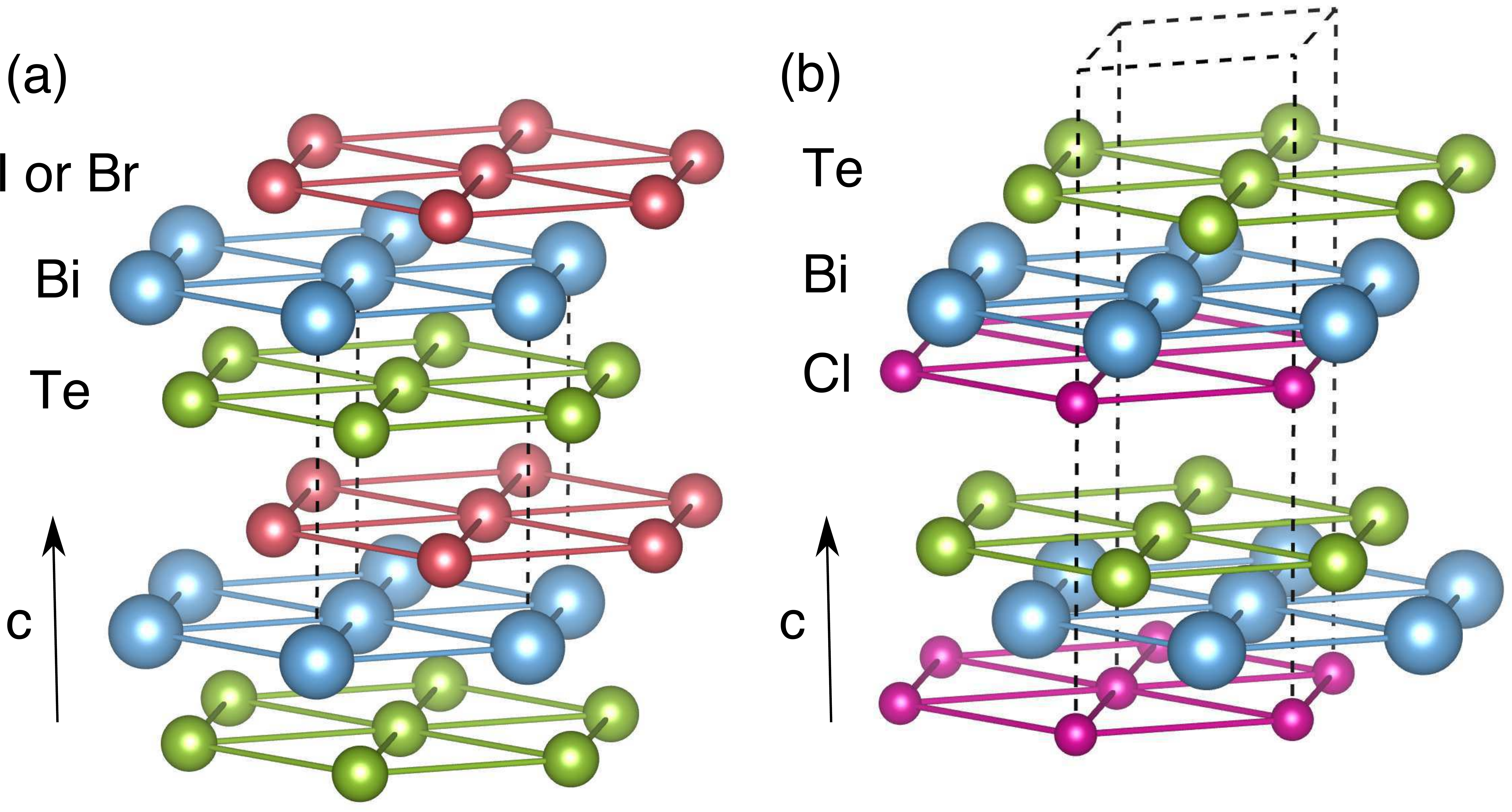}
\caption{Structures of (a) BiTeI and BiTeBr, and (b) BiTeCl. The
layers are stacked along the $c$ direction, and the stacking leads
to a trigonal $P3m1$ space group in BiTeI and BiTeBr, and to a 
hexagonal $P6_3mc$ space group with doubled cell in BiTeCl. The
primitive cells are indicated with dashed lines.}
\label{fig:structures}
\end{figure}

\begin{table*}
  \setlength{\tabcolsep}{2pt} 
  \caption{Equilibrium lattice parameters (in \AA) of the bismuth
  tellurohalides using the LDA, PBE, PBEsol, and PBE+TS exchange
  correlation functionals, and corresponding experimental data
  from Ref.~\onlinecite{bitex_structures}.} 
  \label{tab:latt_params}
  \begin{ruledtabular}
  \begin{tabular}{c|cc@{\hskip 0.6cm}cc@{\hskip 0.6cm}cc@{\hskip 0.6cm}cc@{\hskip 0.6cm}cc}
  &
  \multicolumn{2}{c}{LDA}\hspace{0.6cm} &
  \multicolumn{2}{c}{PBE}\hspace{0.6cm} &
  \multicolumn{2}{c}{PBEsol}\hspace{0.6cm} &
  \multicolumn{2}{c}{PBE$+$TS}\hspace{0.6cm} &
  \multicolumn{2}{c}{Experiment} \\
            &  $a$   & $c$ &  $a$   & $c$ &  $a$   & $c$ &  $a$   & $c$ &  $a$   & $c$ \\
  \hline
  BiTeI  &  $4.31$ & $6.64$  &  $4.43$ & $7.46$ &  $4.34$ & $6.81$ & $4.33$ & $6.63$ & $4.34$ & $6.85$ \\
  BiTeBr &  $4.24$ & $6.28$  & $4.36$ & $7.06$ &  $4.27$ & $6.48$  & $4.43$ & $6.74$ & $4.27$ & $6.49$   \\
  BiTeCl &  $4.22$ & $11.99$ & $4.32$ & $13.61$ &  $4.24$ & $12.50$   & $4.31$ & $12.87$ & $4.24$ & $12.40$  \\
\end{tabular}
\end{ruledtabular}
\end{table*}

\begin{table*}
  \setlength{\tabcolsep}{2pt} 
  \caption{Equilibrium distances (in \AA) between the Bi--Te and Bi--halide 
  planes in the bismuth tellurohalides using the LDA, PBE, PBEsol, 
  and PBE+TS exchange-correlation functionals, and corresponding 
  experimental data from Ref.~\onlinecite{bitex_structures}.}
  \label{tab:internal}
  \begin{ruledtabular}
  \begin{tabular}{c|cc@{\hskip 0.6cm}cc@{\hskip 0.6cm}cc@{\hskip 0.6cm}cc@{\hskip 0.6cm}cc}
  &
  \multicolumn{2}{c}{LDA}\hspace{0.6cm} &
  \multicolumn{2}{c}{PBE}\hspace{0.6cm} &
  \multicolumn{2}{c}{PBEsol}\hspace{0.6cm} &
  \multicolumn{2}{c}{PBE$+$TS}\hspace{0.6cm} &
  \multicolumn{2}{c}{Experiment} \\
            &  Bi--Te  & Bi--halide &  Bi--Te  & Bi--halide &  Bi--Te & Bi--halide &  Bi--Te  & Bi--halide &  Bi--Te   & Bi--halide \\
  \hline
  BiTeI  &  $1.71$ & $2.11$  &  $1.71$ & $2.11$ &  $1.72$ & $2.11$ & $1.78$ & $2.19$ & $2.10$ & $1.72$ \\
  BiTeBr &  $1.74$ & $1.88$  & $1.74$ & $1.88$ &  $1.75$ & $1.88$  & $1.74$ & $1.87$ & $1.81$ & $1.81$   \\
  BiTeCl &  $1.75$ & $1.66$ & $1.76$ & $1.68$ &  $1.76$ & $1.68$   & $1.75$ & $1.67$ & $1.76$ & $1.76$  \\
\end{tabular}
\end{ruledtabular}
\end{table*}

The bismuth tellurohalides are layered materials with a polar structure
as shown in Fig.~\ref{fig:structures}. The structure of BiTeI belongs to the
trigonal $P3m1$ space group, where the stacking direction $c$ is also
the polar axis.~\cite{bitex_structures} Early x-ray powder diffraction
data for BiTeBr revealed a structure similar to that of BiTeI, but with
random disorder in the Te and Br sites.~\cite{bitex_structures} A more
recent study reports an ordered phase of the same symmetry as BiTeI
replacing the I atoms by Br
atoms,~\cite{bitebr_bitecl_rashba_experiment} and in this work we use
the latter ordered phase. The structure of BiTeCl is also ordered, but
the stacking differs from that of BiTeI; the primitive cell is
doubled along the stacking direction, resulting in a structure in the
hexagonal space group $P6_3mc$.~\cite{bitex_structures}

DFT calculations are performed to relax the lattice parameters and
internal coordinates with residual forces below $1$\,meV/\AA\@ and
residual stresses below $0.01$\,GPa. The relaxed lattice parameters are
provided in Table~\ref{tab:latt_params}, and the results have been
obtained using the local density approximation (LDA) to the exchange
correlation
functional,~\cite{PhysRevLett.45.566,PhysRevB.23.5048,PhysRevB.45.13244}
the generalized gradient approximation of Perdew-Burke-Ernzerhof
(PBE),~\cite{PhysRevLett.77.3865} the PBE approximation for solids
(PBEsol),~\cite{pbesol_functional} and the PBE approximation with the
Tkatchenko-Scheffler van der Waals correction (PBE+TS)~\cite{ts_vdW}. 
A comparison with the reported experimental lattice parameters
from Ref.~\onlinecite{bitex_structures} suggests that the cell volumes 
are most accurately captured by the PBEsol functional. 

The only free internal parameters are the distances between the Bi--Te 
and Bi--halide planes, which we show in Table~\ref{tab:internal} for the 
same selection of exchange-correlation functionals.
For BiTeI, it has previously been pointed
out~\cite{bahramy_soc_microscopic_bitei} that the experimental assignment
had the Te and I locations reversed, an error attributed to the
similar x-ray scattering strengths of Te and I.
Taking this into account, the structures obtained using the 
semilocal functionals are in good agreement with the experimental structure.
For BiTeCl, the reported structure is such that the Te and Br sites exhibit 
random disorder, and therefore there is a unique interplanar distance. This 
distance is intermediate between the calculated interplanar distances for
the ordered structure that we use. For BiTeCl, again there is reasonable
agreement between theory and experiment.

Overall, the structural study suggests that PBEsol accurately describes the 
bismuth tellurohalides, and we have therefore used this functional for all
subsequent calculations. We also note that our calculated structural
parameters are in good agreement with previous computational 
reports.~\cite{bahramy_soc_microscopic_bitei}

\subsection{Electronic properties}

The bismuth tellurohalides are small band gap semiconductors exhibiting
strong Rashba splitting in both the valence and conduction bands near
the band gap minimum. This occurs around the $A$-point at the center of
the electronic Brillouin zone on the $k_z=\pi/c$ plane for BiTeI and
BiTeBr, and around the $\Gamma$-point in BiTeCl due to band folding
arising from the doubling of the primitive cell along the stacking
direction $c$. Experimental samples exhibit $n$-doping, which makes the
Rashba splitting on the conduction bands the most relevant for potential
applications.

\subsection{Lattice dynamics}

The effects of temperature on the band structure can be divided into
contributions from electron-phonon coupling and from thermal expansion.
As a starting point to calculate both contributions from first
principles, lattice dynamics calculations have to be performed. We use
the finite displacement approach~\cite{phonon_finite_displacement} to
lattice dynamics in conjunction with nondiagonal
supercells.~\cite{non_diagonal} For BiTeI, we find that vibrational
energies calculated from the Fourier interpolation over fine grids
starting with coarse grids of sizes $4\times4\times4$ and
$6\times6\times6$ $\mathbf{q}$-points are very similar. We therefore
report results using a coarse $4\times4\times4$ $\mathbf{q}$-point grid
for BiTeI and BiTeBr, and a coarse $4\times4\times2$ $\mathbf{q}$-point
grid for BiTeCl. All calculations include the spin-orbit interaction,
which has been found to affect the vibrational frequencies
of the bismuth tellurohalides significantly.~\cite{bitei_lattice_dynamics_chulkov}

\subsection{Electron-phonon coupling} \label{subsec:elph}

The electron-phonon coupling contribution to the temperature dependence
of an electronic eigenvalue $\epsilon_{n\mathbf{k}}$, labeled by
quantum numbers $(n,\mathbf{k})$, is determined within the adiabatic
approximation by its expectation value with respect to the vibrational
density
\begin{equation}
\epsilon_{n\mathbf{k}}(T)=\frac{1}{\mathcal{Z}}
   \sum_{\mathbf{s}}\langle\Phi_{\mathbf{s}}(\mathbf{u})|
   \epsilon_{n\mathbf{k}}(\mathbf{u})|\Phi_{\mathbf{s}}
   (\mathbf{u})\rangle e^{-E_{\mathbf{s}}/k_{\mathrm{B}}T}. \label{eq:elph}
\end{equation}
In this equation, the vibrational wave function
$|\Phi_{\mathbf{s}}(\mathbf{u})\rangle$ in state $\mathbf{s}$ has energy
$E_{\mathbf{s}}$ and is described within the harmonic approximation,
$\mathbf{u}=\{u_{\nu\mathbf{q}}\}$ is a collective coordinate for all
the nuclei written in terms of normal modes of vibration
$(\nu,\mathbf{q})$,
$\mathcal{Z}=\sum_{\mathbf{s}}e^{-E_{\mathbf{s}}/k_{\mathrm{B}}T}$ is
the partition function, $T$ is the temperature, and $k_{\mathrm{B}}$ is
Boltzmann's constant.

We evaluate Eq.~(\ref{eq:elph}) using a quadratic approximation to
$\epsilon_{n\mathbf{k}}(\mathbf{u})$ in terms of normal
coordinates,~\cite{0022-3719-9-12-013,PhysRevLett.105.265501,
  monserrat_elph_diamond_silicon,jcp_ponce_convergence}
as well as using a Monte Carlo integration
technique.~\cite{giustino_nat_comm,thermal_lines} 
The second order expansion coefficients appearing in the quadratic
approximation for polar modes diverge in the limit
$\mathbf{q}\to0$,~\cite{jcp_ponce_convergence} but in practice we find
that by choosing a fixed amplitude of $0.5\sqrt{\langle
u_{\nu\mathbf{q}}^2\rangle}$ in a finite differences context leads to
good agreement with the Monte Carlo method, which does not exhibit any
divergent behavior. Because nondiagonal supercells can be used in
conjunction with the quadratic approximation to efficiently sample the
vibrational Brillouin zone,~\cite{non_diagonal} we choose this method
for the electron-phonon coupling calculations reported below. We also
find that the results do not change significantly between grid sizes of
$4\times4\times4$ and $6\times6\times6$ in BiTeI. We therefore use
$4\times4\times4$ grids for BiTeI and BiTeBr, and $4\times4\times2$
grids for BiTeCl.

\subsection{Thermal expansion} \label{subsec:th_expansion}

We evaluate the effects of thermal expansion within the quasiharmonic
approximation.~\cite{dove_lattice_dynamics_book} We calculate the
vibrational free energy at temperature $T$ as a function of the lattice
parameters $a$ and $c$, and minimize the total Gibbs free energy at each
temperature with respect to the value of the lattice parameters to
determine the equilibrium volume at that temperature. Due to the layered
nature of the bismuth tellurohalides, the thermal expansion along the
stacking $c$ direction is about an order of magnitude stronger than that
along the in-plane direction, and therefore we minimize the Gibbs free
energy independently for each lattice parameter to reduce the number of
calculations required.

\section{Temperature dependence of the Rashba splitting}

\subsection{Temperature dependent Rashba splitting in BiTeI, BiTeBr, and BiTeCl} \label{subsec:bulk_bitex}

\begin{figure}
\centering
\includegraphics[scale=0.31]{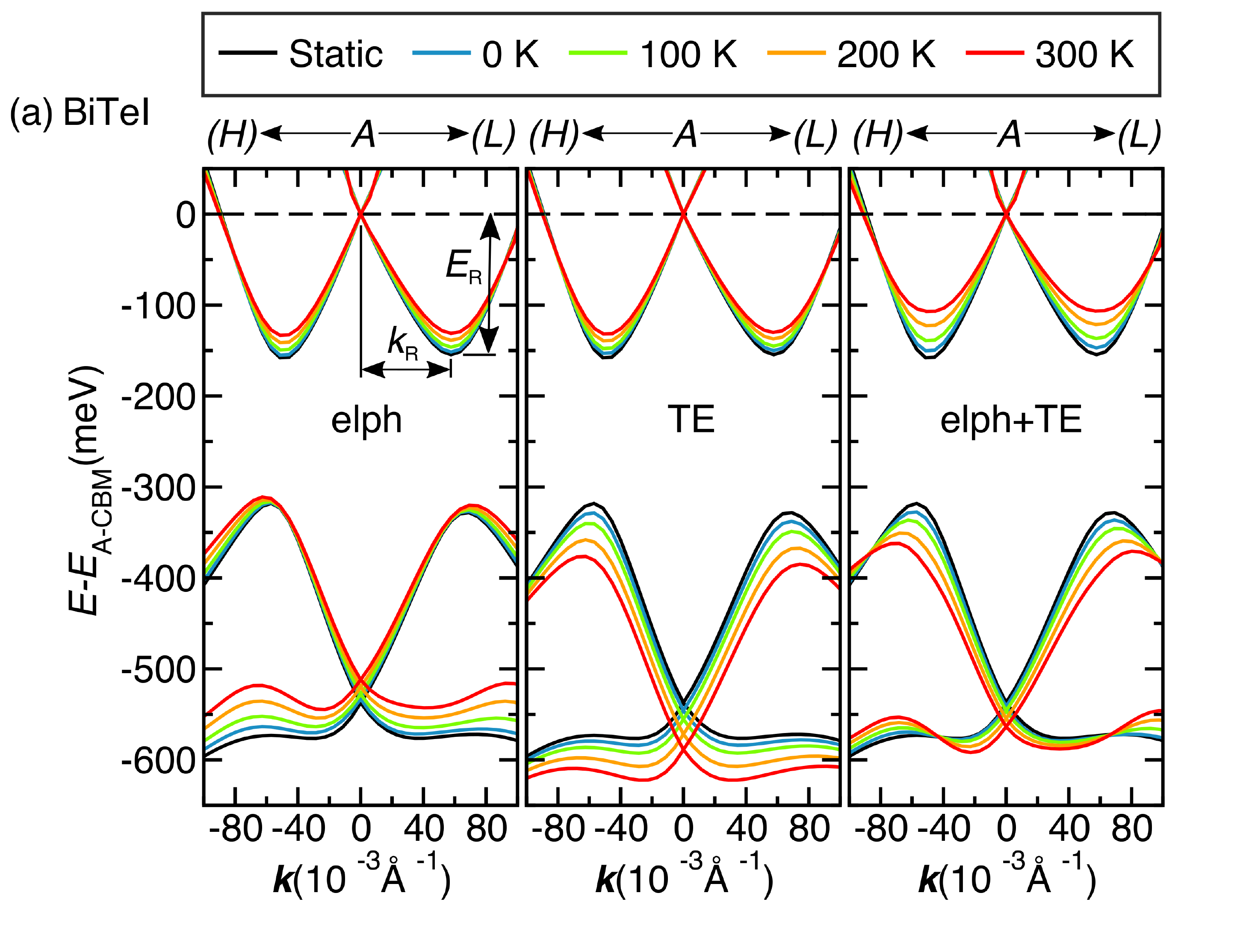}
\includegraphics[scale=0.31]{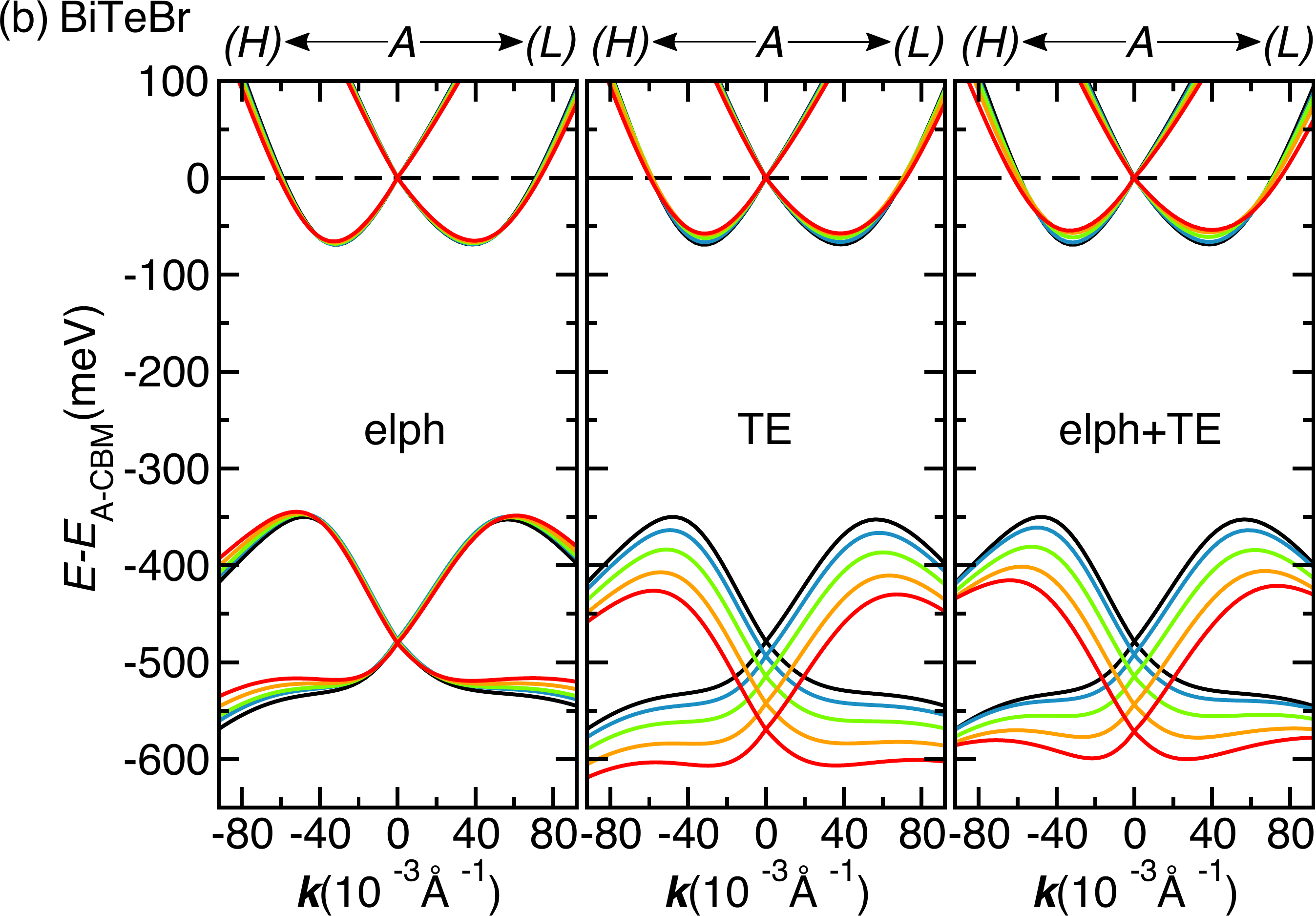}
\includegraphics[scale=0.31]{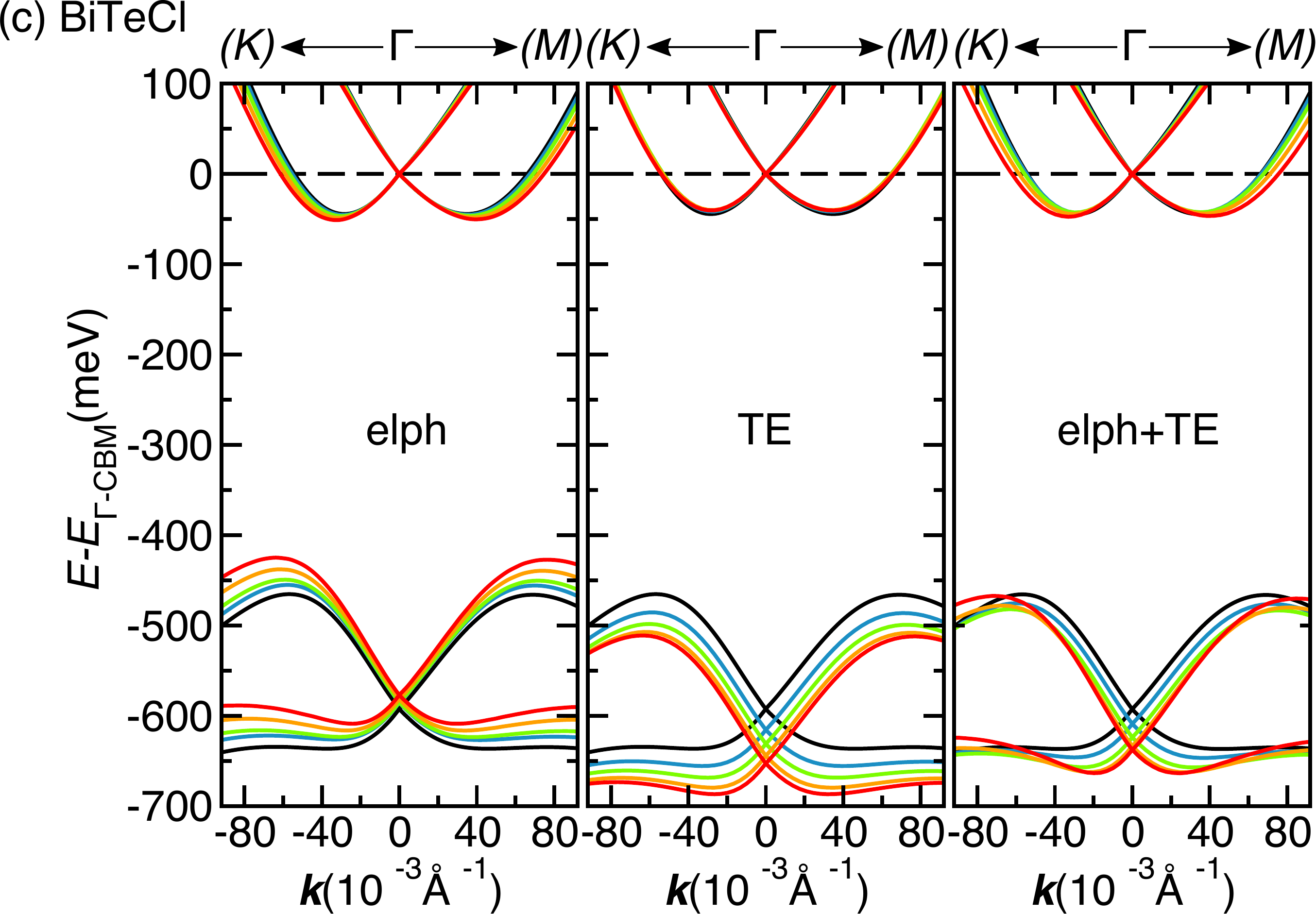}
\caption{Temperature dependence of the Rashba-split bands of
(a) BiTeI, (b) BiTeBr, and (c) BiTeCl.
Contributions from electron-phonon coupling (elph) and thermal-expansion
(TE) are combined to obtain the overall temperature dependence
(elph+TE) in left, middle, and right panels respectively.  The
electronic states are referenced to the conduction-band minimum
at the $A$-point for BiTeI and BiTeBr and the $\Gamma$-point
for BiTeCl. Note that the plotted regions only represent about
$10$\% of the distance from the $A$ or $\Gamma$ points to the
indicated zone boundaries.  The definitions of $E_{\mathrm{R}}$
and $k_{\mathrm{R}}$ are depicted in the upper-left panel.}
\label{fig:tdep-BTI}
\end{figure}

The temperature dependence of the Rashba-split conduction and valence
bands of BiTeI, BiTeBr, and BiTeCl is shown in Fig.~\ref{fig:tdep-BTI}.
The contributions from electron-phonon coupling and thermal expansion are
plotted separately, showing that their effect is similar in magnitude.
The combination of both effects leads to the overall temperature
dependence. The bands are referenced with respect to the lowest 
conduction-band state at the $A$-point in BiTeI and BiTeBr, and to 
the lowest conduction-band state at the $\Gamma$-point in BiTeCl.
We also emphasize that the 
$\mathbf{k}$-space region shown in Fig.~\ref{fig:tdep-BTI} only covers
about 10\% of the Brillouin-zone dimensions.


\begin{figure}
\centering
\includegraphics[scale=0.33]{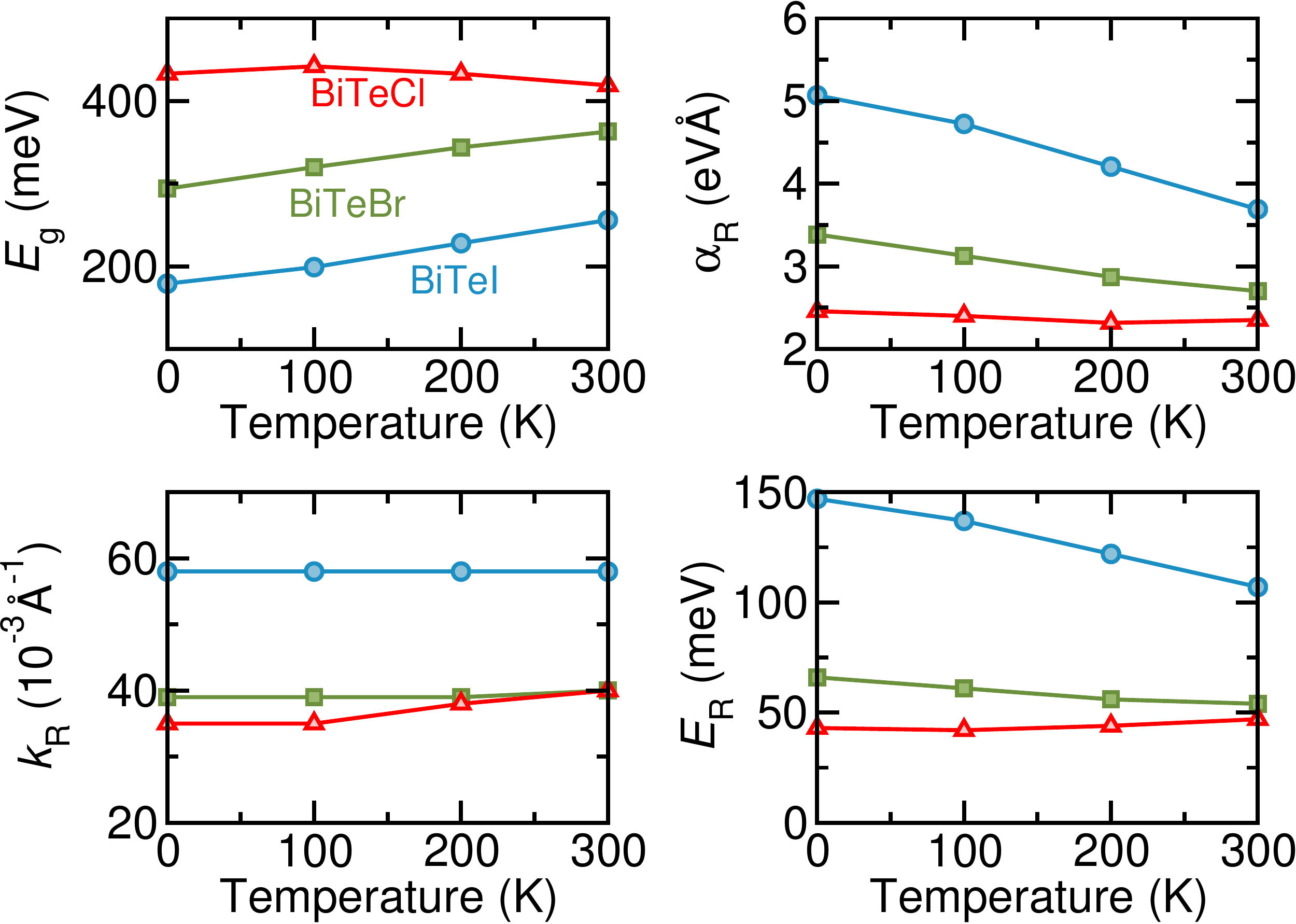}
\caption{Temperature dependence of the minimum band gap and of the
conduction-band Rashba parameters of the bismuth tellurohalides.}
\label{fig:rashba}
\end{figure}

In all materials, the band dispersion deviates significantly from
that of the two-band model of Eq.~(\ref{eq:rashba}).
Nonetheless, it is still common practice to extract Rashba parameters
from these band structures, and in the following we will determine
$k_{\mathrm{R}}$ and $E_{\mathrm{R}}$ from the dispersion
along the $A$--$L$ line for BiTeI
and BiTeBr, and along the $\Gamma$--$M$ line for BiTeCl, as indicated in
Fig.~\ref{fig:tdep-BTI} for BiTeI. We show a summary of the temperature
dependence of the minimum band gap and of the Rashba parameters in 
Fig.~\ref{fig:rashba}.

For BiTeI, the $E_{\mathrm{R}}$ parameter of the conduction bands
decreases with increasing temperature due to both electron-phonon
coupling and thermal expansion.
In contrast, the $k_{\mathrm{R}}$ parameter is nearly
temperature independent. The $E_{\mathrm{R}}$ parameter of
the valence bands shows a weak temperature dependence, while both
electron-phonon coupling and thermal expansion lead to an increase in
the $k_{\mathrm{R}}$ parameter with increasing temperature. The 
minimum band gap size increases with increasing temperature. The Rashba
parameter of the conduction bands at $0$\,K
of $\alpha_{\mathrm{R}}=5.07$\,eV\AA\@ is somewhat smaller than the static
lattice parameter value of $\alpha_{\mathrm{R}}=5.35$\,eV\AA\@, a
consequence of quantum zero-point motion. With increasing temperature,
the Rashba parameter decreases to $\alpha_{\mathrm{R}}=3.69$\,eV\AA\@ at
$300$\,K.

For BiTeBr, both electron-phonon coupling and thermal-expansion
contributions exhibit the same trends as in BiTeI. However, the
electron-phonon coupling contribution is significantly weaker, such that
the overall temperature dependence of the Rashba-split bands is
dominated by thermal expansion. As in BiTeI, the minimum band gap size increases
with increasing temperature, and the Rashba parameter of the conduction
bands decreases with increasing temperature. The static Rashba parameter
of $\alpha_{\mathrm{R}}=3.63$\,eV\AA\@ decreases to
$\alpha_{\mathrm{R}}=3.39$\,eV\AA\@ at $0$\,K due to quantum zero-point
motion, and to $\alpha_{\mathrm{R}}=2.70$\,eV\AA\@ at $300$\,K due to
thermal motion.

For BiTeCl, the electron-phonon contribution is opposite to
that calculated in BiTeI and BiTeBr, and with increasing temperature it
leads to a reduction of the minimium band gap size and to an increase of the
strength of Rashba splitting of the conduction bands. The distinct
effects of electron-phonon coupling can be attributed to the larger band
gap of BiTeCl compared to that of the heavier bismuth tellurohalides,
which suppresses the interband transitions in favor of the intraband
transitions compared to the compounds with smaller band
gap.~\cite{elph_topological_prb} The thermal-expansion
contribution in BiTeCl exhibits the same trend as that of BiTeI and
BiTeBr, but the temperature dependence is weaker. Overall, the minimum band gap
of BiTeCl has a weak temperature dependence, and the Rashba parameter of
the conduction band decreases by about $0.01$\,eV\AA\@ from $0$\,K to
$300$\,K, determined by a small increase of the Rashba energy which is
overcompensated by a small increase of the Rashba momentum.


We next discuss the origin of the decrease in Rashba splitting with
increasing temperature predicted in the bismuth tellurohalides, and we
focus on BiTeI which exhibits the strongest changes. Following Bahramy
and co-workers, the spin-orbit interaction driving Rashba splitting can
be treated within perturbation theory in a $\mathbf{k\cdot p}$ model,
with a leading-order correction to an electronic state
$\epsilon_{n\mathbf{k}}$ given by~\cite{bahramy_soc_microscopic_bitei}
\begin{equation}
\Delta \epsilon_{n\mathbf{k}}=\sum_{m\neq n} \frac{\langle m\mathbf{k}_0|H_{\mathrm{SOC}}|
  n\mathbf{k}_0\rangle\langle n\mathbf{k}_0|\Delta\mathbf{k\cdot p}|m\mathbf{k}_0\rangle}
  {\epsilon_{n\mathbf{k}_0}-\epsilon_{m\mathbf{k}_0}}+h.c.,
  \label{eq:so_perturb}
\end{equation}
where $\Delta\mathbf{k=k-k}_0$ is the crystal momentum measured from
a reference $\mathbf{k}_0$ taken to be the $A$ point, $\mathbf{p}$ 
is the orbital 
momentum, and $H_{\mathrm{SOC}}$ is the spin-orbit Hamiltonian. 
Equation~(\ref{eq:so_perturb}) allows us to distinguish three contributions 
to the strength of Rashba splitting:~\cite{bahramy_soc_microscopic_bitei}
(i) the magnitude of spin-orbit coupling encoded by $H_{\mathrm{SOC}}$,
(ii) the selection rules on the matrix elements in the numerator, and
(iii) the energy difference in the denominator.
Contributions (i) and (ii) are well optimized in the bismuth
tellurohalides as they
are comprised of heavy elements exhibiting strong spin-orbit coupling.
They also have crystal-field splittings of the valence and conduction bands of
opposite sign, making the two bands symmetry-equivalent and therefore
allowing the relevant matrix elements. Bahramy and co-workers argued
that contribution (iii) is also optimal in the bismuth tellurohalides, as the
small $A$-point band gaps ($\Gamma$-point 
for BiTeCl) make the denominators in Eq.~(\ref{eq:so_perturb}) small, 
enhancing the Rashba splitting.~\cite{bahramy_soc_microscopic_bitei}

In our analysis, the energy denominator in Eq.~(\ref{eq:so_perturb})
becomes temperature-dependent and therefore dominates the temperature
dependence of the Rashba splitting. Thermal expansion increases the
interatomic distances and drives the system towards the atomic limit.
This leads to an increase in the $A$-point band gap
which therefore reduces the
Rasbha splitting according to Eq.~(\ref{eq:so_perturb}), and in
agreement with our first-principles calculations. The electron-phonon
contribution is harder to interpret: different phonon modes couple
differently to electronic states, and although the band gap at the
$A$-point decreases with increasing temperature, the minimum band gap
increases with increasing temperature. The Bi $p_z$-like states
dominating the conduction band and the Te and I $p_z$-like states
dominating the valence band are oriented along the stacking
$c$-direction in the bismuth tellurohalides (see
Fig.~\ref{fig:structures}). As a consequence, the phonon modes that
dominate electron-phonon coupling are those that correspond to atomic
vibrations that change the interlayer distance.


\subsection{BiTeI in the topological-insulator phase} \label{subsec:ti}

Our results indicate that the Rashba splitting decreases with increasing
temperature in the bismuth tellurohalides, driven by both
electron-phonon coupling and thermal expansion. The thermal-expansion
contribution is determined by the approach to the atomic limit and the
associated band-gap increase, as discussed in the previous section. This
raises the possibility that in systems with inverted bands, the Rashba
splitting would increase with increasing temperature. It has been
theoretically predicted that BiTeI undergoes a band inversion and
becomes a topological insulator under hydrostatic
pressure,~\cite{bahramy_bitei_ti} with the topological phase transition
mediated by a Weyl semimetal
phase.~\cite{murakami_weyl_semimetal,liu_noncentrosymm_weyl}
Experimental efforts to confirm this prediction have so far reached
contradictory conclusions, with a topological phase transition reported
around $2$--$4$\,GPa in
Refs.~\onlinecite{carr_bitei_ti_experiment,yanming_ma_bitei_pressure,
  tokura_bitei_ti_experiment,kim_bitei_ti_experiment,felser_bitei_ti_experiment}
and no topological phase transition reported in
Ref.~\onlinecite{akrap_bitei_ti_experiment}. Similar discussions exist
for BiTeBr and
BiTeCl.~\cite{sasagawa_bitecl_ti_experiment,chulkov_bitebr_ti_theory,
  akrap_bitebr_ti_experiment,sasagawa_bitebr_ti_experiment}
In our calculations we observe a pressure-induced topological phase
transition in BiTeI between $1$--$2$\,GPa.

\begin{figure}
\centering
\includegraphics[scale=0.32]{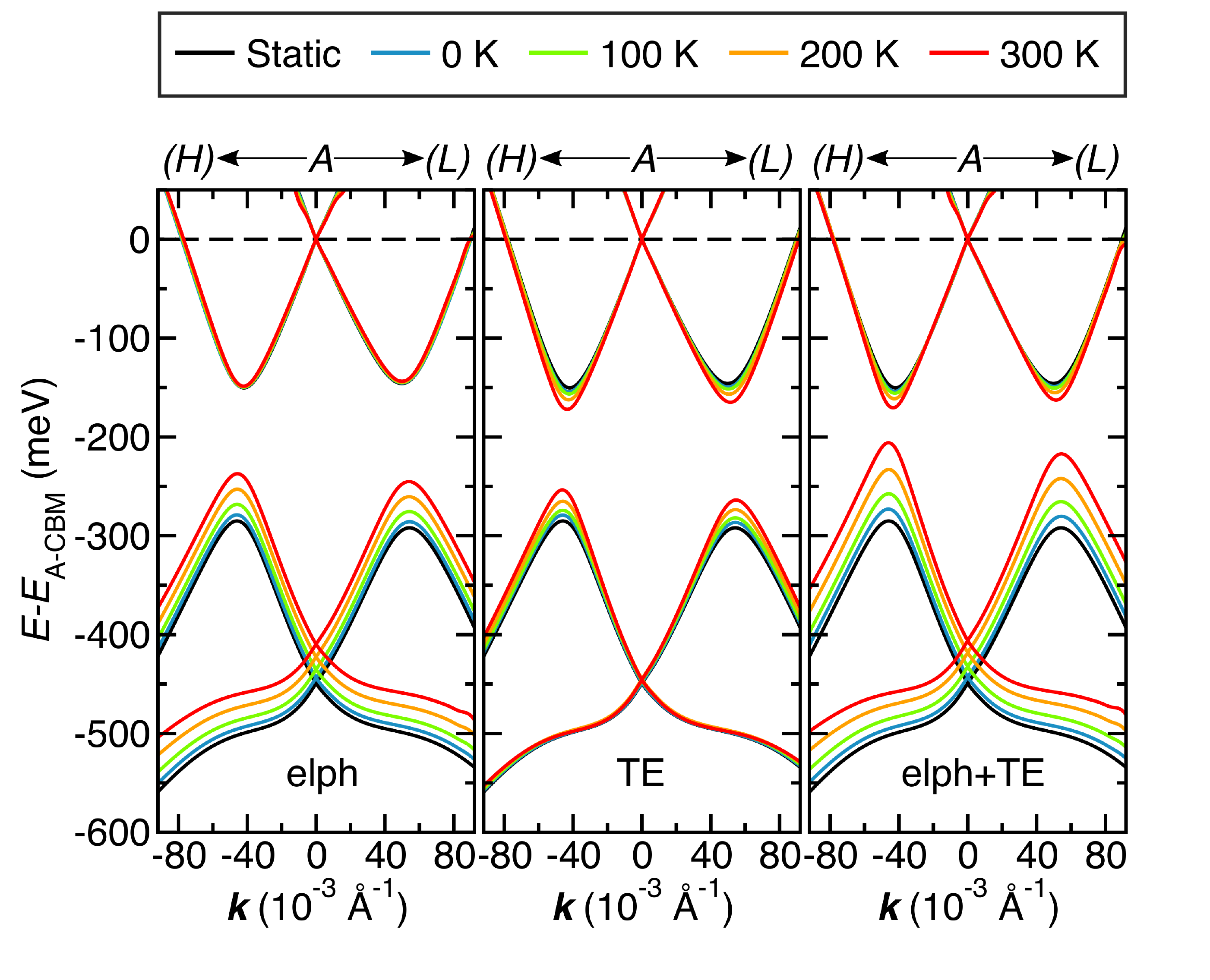}
\caption{Temperature dependence of the Rashba-split bands of BiTeI
in the topological-insulator phase at $4$\,GPa. Contributions from
electron-phonon coupling (elph) and thermal expansion (TE)
are combined to obtain the overall temperature
dependence (elph$+$TE) in left, middle, and right panels respectively. 
The electronic states are referenced
to the conduction-band minimum at the $A$-point. We note that
the plotted regions only represent about
$10$\% of the distance from the $A$ point to the indicated zone
boundaries.}
\label{fig:tdep-BTI-4GPa}
\end{figure}

Our results at $4$\,GPa in the topological phase
are presented in Fig.~\ref{fig:tdep-BTI-4GPa}, where again
we show the temperature dependence of the Rashba-split bands near the
$A$-point of the electronic Brillouin zone. Electron-phonon coupling
provides an almost rigid shift of the Rashba-split bands, which leads
to a decrease of the minimum and $A$-point band gaps 
with increasing temperature. As in the normal phase, the electron-phonon
contribution is complex and depends on multiple phonon modes, which 
overall lead to a small change in the Rashba parameters.
As expected, thermal expansion
leads to a reduction of the band gap and to an increase of the Rashba
splitting with increasing temperature. Overall, the Rashba splitting
increases with increasing temperature for both the conduction and
valence bands.

In view of these results, we propose a new signature for identifying the
topological phase transition in BiTeI. A measurement of the temperature
dependence of the minimum band gap or the $A$-point band gap
should show an increase of the band-gap size
at pressures in which BiTeI is a normal 
insulator, and to a decrease of the band-gap size at pressure in which
BiTeI is a topological insulator. Monitoring the temperature dependence
of the band gap could be a general approach for identifying
pressure-induced topological phase transitions.~\cite{monserrat_ti_temp}

We also note that the temperature dependent spin-split bands shown in
Fig.~\ref{fig:tdep-BTI-4GPa} deviate significantly from the canonical 
dispersion that arises from the Rashba Hamiltonian in Eq.~(\ref{eq:rashba}).
At the topological phase transition, the bands cross linearly in a cone-like
fashion. In the regime shown in Fig.~\ref{fig:tdep-BTI-4GPa}, which is close 
to the topological phase transition, the bands can be described by two cone-like structures which exhibit avoided crossings and therefore deviate significantly 
from a picture of parabolic intersecting Rashba bands. A similar behavior was 
observed when studying the pressure-induced topological phase transition 
in BiTeI.~\cite{bahramy_bitei_ti}

\subsection{Discussion} \label{subsec:exp}

There are several limitations to our analysis. The first is the
description of the electronic degrees of freedom within semilocal DFT.
Many-body electron correlations modify the band-gap size and the Rashba
splitting strength of the bismuth tellurohalides,~\cite{bitei_many_body}
and a full treatment would require the simultaneous inclusion of
electron correlations and temperature. Nonetheless, we expect that the
general trends presented in this work would
remain.~\cite{gonze_gw_elph,gw_thermal_lines} The second limitation
concerns the neglect of nonadiabatic contributions, which are typically
small but could become larger when the band gaps become small. Overall,
due to the band-gap underestimation we obtain using semilocal DFT, we
expect that our results represent an upper bound on the real temperature
dependence of the Rashba splitting of the bismuth tellurohalides.
Nonetheless, the temperature dependence that we predict should still be
observable experimentally, and we suggest that tuning the band gap to
smaller sizes, for example by application of hydrostatic pressure, could
enhance the temperature dependence of the Rashba splitting and
facilitate observation.

\begin{figure}
\centering
\includegraphics[scale=0.36]{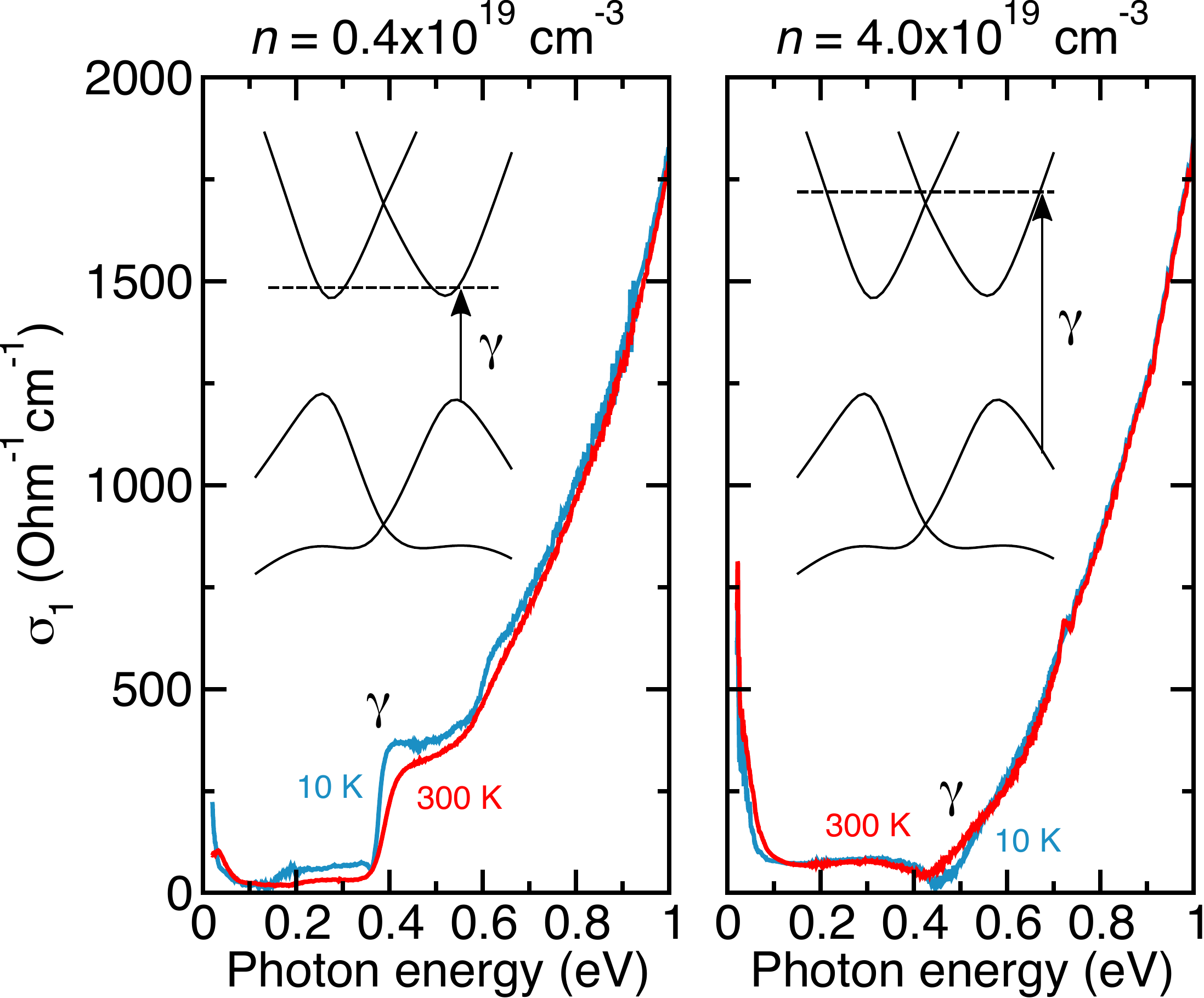}
\caption{Experimental optical conductivity $\sigma_1$ of BiTeI at an electron
doping concentration of $n=0.4\times10^{19}$\,cm$^{-3}$ (left) and
$n=4.0\times10^{19}$\,cm$^{-3}$ (right) at $10$\,K (blue curves) and
at $300$\,K (red curves). The inset diagrams depict the position of
the Fermi level (dashed line) within the Rashba-split conduction bands, and the
relevant interband $\gamma$ transition. The experimental data was
provided by Jongseok Lee and is partially published in
Ref.~\onlinecite{bitei_optical_exp_theory}.}
\label{fig:experiment}
\end{figure}

To test our predictions, we compare to
the optical conductivity spectrum of BiTeI
measured by Lee and co-workers,~\cite{bitei_optical_exp_theory} and shown 
in Fig.~\ref{fig:experiment}. It is 
important to note that experimental samples are typically $n$-doped, and the
data shown in Fig.~\ref{fig:experiment} corresponds to two different samples
with doping concentrations of $n=0.4\times10^{19}$\,cm$^{-3}$ with the Fermi
level near the bottom of the conduction band (left diagram), and of 
$n=4.0\times10^{19}$\,cm$^{-3}$ with the Fermi level near the crossing of the
Rashba-split bands (right diagram).~\cite{bitei_optical_exp_theory} In both
cases we focus on the band gap determined by the interband $\gamma$ 
transition, which is indicated in 
the insets of Fig.~\ref{fig:experiment}, although we note that intraband
transitions between the spin-split conduction bands are also 
observed.~\cite{bitei_optical_exp_theory,bitei_optical_exp_arxiv} 
For the weakly doped sample, the size of the band gap increases with increasing
temperature (the position of the $\gamma$ edge shifts to higher energies).
This is consistent with our prediction of the increase of the band gap with
increasing temperature for BiTeI. In the more highly doped sample, the 
position of
the $\gamma$ edge exhibits the opposite behavior, and shifts to lower 
energies with increasing temperature. In this latter sample, the Fermi level
sits around the crossing point between the spin-split bands. Assuming 
that the only effects of doping are a rigid shift of the Fermi level, then 
our results for BiTeI also agree with this observation. This is because,
as shown in Fig.~\ref{fig:tdep-BTI}, for electronic 
states of wavevectors at distances larger than about $80\times10^{-3}$\,\AA$^{-1}$
from the $A$-point, the energy gap decreases with increasing temperature.
This comparison confirms our predictions, and highlights the complex interplay
between the level of doping and the temperature dependence of the band
structure as measured by optical probes. We note that the temperature
dependence of the band gap of BiTeBr and BiTeCl is reported in 
Refs.~\onlinecite{tdep_exp_BTB_BTC,tdep_exp_BTC_tanner},
where a decrease of the band gap with increasing temperature is observed, 
and consistently with our discussion for BiTeI, in all cases the electron 
doping level is significant. 


An alternative experimental test of our predictions would be to use
angle-resolved photoemission spectroscopy (ARPES), which could
provide a direct measurement of the temperature-induced changes
to the Rashba-split bands. However, using this technique it might be
difficult to disentangle the bulk and surface
dispersions.~\cite{bitei_giant_rashba,grioni_rashba_bitei_valence_conduction,
  bitei_disentanglement_bulk_surface}
In this context, another consequence of our results is to question 
the use of standard
first-principles calculations to interpret ARPES experiments in this
class of materials. The original report of giant Rashba splitting in
BiTeI was based on ARPES measurements.~\cite{bitei_giant_rashba}
One of the arguments in support of the bulk nature of this Rashba
splitting was the good agreement of the experimental Rashba parameters
with first-principles calculations based on a generalized gradient
approximation (GGA) to the exchange-correlation functional and performed
at zero temperature. Subsequent experiments questioned the bulk nature
of the Rashba splitting reported in
Ref.~\onlinecite{bitei_giant_rashba}, as ultraviolet ARPES probing the
surface bands led to a dispersion in good agreement with that of
Ref.~\onlinecite{bitei_giant_rashba}, but soft x-ray ARPES probing the
bulk bands showed that the surface Rashba parameter was $20$\% larger
than the bulk parameter.~\cite{bitei_disentanglement_bulk_surface}
Again, part of the justification for these results was the support of
first-principles GGA calculations. Our results show the
changes in the Rashba parameters induced by temperature are comparable
to the measured difference between surface and bulk Rashba splittings.
Together with previously reported contributions from electron
correlation,~\cite{bitei_many_body} which are similar in size to those
from temperature effects,
our results show that predictions using first-principles
semilocal DFT methods within the static lattice approximation have to be
interpreted with caution.


More generally, our work contributes to the investigations of the
interplay between temperature and spin-orbit physics. Recent studies
have reported significant effects of temperature on the properties of
topological insulators and Weyl semimetals, including
temperature-induced topological phase
transitions.~\cite{elph_topological_prl,elph_topological_prb,
  elph_topological_jhi,garate_ph_linewidths_band_inversion,
  monserrat_ti_temp,antonius_ti_temp,duane_phonon_weyl}
Our calculations
show that temperature can also significantly renormalize the Rashba
splitting, suggesting that finite temperatures can dramatically modify
spin-orbit physics in a variety of contexts.

\section{Summary} \label{sec:summary}

We have performed first-principles calculations to study the temperature
dependence of the Rashba splitting in the bismuth tellurohalides. We
find a reduction in the Rashba splitting with increasing temperature
which is particularly strong in BiTeI, with a $40$\% change in going
from $0$\,K to $300$\,K. We find the opposite behavior when BiTeI has
inverted bands in the topological-insulator phase, with temperature
enhancing the Rashba splitting, and propose this reversal of the
temperature dependency of the band gap as a signature for identifying a
pressure-induced topological phase transition. Electron-phonon coupling
and thermal expansion contribute similarly to the temperature dependence
of the Rashba-split bands, and their microscopic behavior is dominated
by the layered nature of the bismuth tellurohalide structures.

Overall, our results show that quantitative first-principles predictions
of Rashba splitting must incorporate the effects of temperature.
Furthermore, the temperature-induced changes that we predict in BiTeI
are consistent with experimental optical conductivities, and additional 
measurements could provide further insights into the
nature of Rashba splitting and topology in these materials.

\acknowledgments

We thank Jongseok Lee for sharing the optical conductivity data, and
Janice Musfeldt, Sang-Wook Cheong, and David Tanner for useful
conversations and correspondence. This work was funded by NSF grant 
DMR-1408838. 
B.M. thanks Robinson College, Cambridge, and the Cambridge Philosophical 
Society for a Henslow Research Fellowship.


\bibliography{rashba}

\end{document}